\documentclass[12pt]{article}
\usepackage[bookmarks,bookmarksnumbered]{hyperref}
\usepackage{graphicx,color}
\usepackage{amsmath,XoohmE}

\let\bs=\boldsymbol
\def\cA{\mathcal{A}}
\def\sA{\mathscr{A}}
\def\bA{\mathbf{A}}
\def\bB{\mathbf{B}}
\def\sB{\mathscr{B}}

\def\Dh{\skew3\widehat{D}}
\def\Db{\relax\leavevmode\hbox{$D$\kern-.6em
               \vrule height1.9ex depth-1.8ex width5pt}\kern2pt}
\def\Dbh{\skew3\widehat{\Db}}
\def\dvd{\mathop{\smash{\rlap{\raisebox{1.7ex}{$\SSS\leftrightarrow$}}{\partial}}}\nolimits}

\def\sF{\mathscr{F}}
\def\fb{\bar\phi}
\def\Fb{\bar{F}}
\def\BFb{{\bar{\bs{\Phi}}}}
\def\BF{\bs{\Phi}}
\def\jb{\bar\psi}
\def\BK{\mathbf{K}}
\def\cK{\mathcal{K}}
\def\sK{\mathscr{K}}
\def\El{{\ell}}
\def\cBL{\bs{\mathcal{L}}}
\def\SL{\mathscr{L}}
\def\lb{{\lambda}}
\def\bl{{\mkern3mu\vrule height1ex depth-.9ex width4pt
          \mkern-9.25mu\lambda}}
\def\Lb{{L}}
\def\BL{{\boldsymbol{\Lambda}}}

\def\Qb{{\bar Q}}
\def\vsb{\bar{\varsigma}}
\def\bX{{\bf X}}

\def\bY{\boldsymbol{\Upsilon}}
\def\pp{{\mathchar'75\mkern-9mu|\mkern3mu}}
\def\mm{{=}}

\def\[{\raisebox{2pt}{,}\mkern1mu}
\def\sm#1{\left[\begin{smallmatrix}#1\end{smallmatrix}\right]}

 \SfTitles
 \allowdisplaybreaks
 \seceq
\begin{document}

\begin{flushright}
\today
\end{flushright}

\begin{center}
{\LARGE\sf\bfseries Worldsheet Matter Superfields on Half-Shell}\\[3mm]
{\sf\bfseries T.\,H\"{u}bsch$^*$ {\rm and} I.E.\,Petrov$^{\dagger\ddagger}$}\\[1mm]
{\small\it
  $^*$Dept.\ of Physics and Astronomy,
  Howard University, Washington, DC 20059, USA
   \\[-1mm]
  $^\dagger$Imaging Research Laboratories, Robarts Research Institute, London, Ontario N6A 5K8, Canada
 }\\[3mm]
{\small\sf\bfseries ABSTRACT}\\[2mm]
\parbox{130mm}{\parindent=2pc\addtolength{\baselineskip}{-2pt}\noindent
In this paper we discuss some of the effects of using ``unidexterous'' worldsheet superfields, which satisfy worldsheet differential constraints $\vd_\mm\BL=0=\vd_\pp\bY$ and so are partly on-shell, \ie, on half-shell.
 Most notably, this results in a stratification of the field space that reminds of ``brane-world'' geometries.
 Linear dependence on such superfields provides a worldsheet generalization of the super-Zeeman effect.
 In turn, non-linear dependence yields additional left-right asymmetric {\em\/dynamical\/} constraints on the propagating fields, again in a stratified fashion.
}
\end{center}
\noindent PACS: 11.30.Pb, 12.60.Jv

\section{Introduction, Results and Synopsis}
 \label{s:IRS}
Supersymmetry has been studied for almost four decades\cite{r1001,rPW,rWB,rBK}, but still seems to harbor novel features. In particular, worldsheet supersymmetry is unique among supersymmetric field theories in that the Lorentz group is abelian $\textsl{Spin}(1,1)\simeq\textsl{U}(1)$. This permits the definition of the {\em\/twisted\/}-chiral superfields\cite{rGHR}, and a whole host of supersymmetry representations not possible in higher-dimensional spacetimes\cite{rHSS}; see \Eqs{e:L}{e:R} for two that we will need herein: the {\em\/lefton\/} and {\em\/rightons\/} superfields, denoted $\BL$ and $\bY$, respectively.

In superfield formulations of supersymmetry it is not uncommon that some of the component fields, called {\em\/auxiliary\/}, end up having algebraic equations of motion. It is routinely assumed that such algebraic equations may be solved for, and their solutions substituted back into the Lagrangians, thereby obtaining an equivalent but simpler description of the model.

It is our main purpose to point out that this is not as straightforward as it may seem, that it may well impose rather non-standard {\em\/dynamical\/} constraints on the model, and may well result in {\em\/stratifying\/} the field space of the model into regions (strata) of varying dimensions and dynamics. As worldsheet models are most often used in superstring theory and suitable subsets of the field space are identified as the effective, ``real'' spacetime, worldsheet models with such stratified field space would seem to provide a natural Lagrangian framework for ``brane-world''-like geometries.

The remainder of this section offers a basic review of $(1,1|2,2)$-superspace notation, with a few technical details deferred to Appendix~\ref{s:D}.
Section~\ref{s:MO} employs this to derive some simple consequences of using lefton superfields $\BL$, and section~\ref{s:SX} showcases these results in a few simple but nontrivial examples.
Section~\ref{s:C} collects our conclusions.

\subsection{$(1,1|2,2)$-Superspace Notation}
 \label{ss:Basic}
Adopting the notation of Refs.\cite{rHSS,rGSS,rIPd}
we list here a few basic definitions and results in $(1,1|2,2)$-superspace notation, wherein the supersymmetry charges $Q_\pm$ and $\Qb_\pm$, superderivatives $D_\pm,\Db_\pm$ and light-cone worldsheet derivatives $\vd_\pp,\vd_\mm$ satisfy:
\begin{subequations}
 \label{e:SuSy}
\begin{alignat}{3}
  \big\{\, Q_- \,,\, \Qb_- \,\big\}&=2i\vd_\mm, &\qquad
  \big\{\, Q_+ \,,\, \Qb_+ \,\big\}&=2i\vd_\pp,  \label{eSusyQ}\\
  \big\{\, D_- \,,\, \Db_- \,\big\}&=2i\vd_\mm, &\qquad
  \big\{\, D_+ \,,\, \Db_+ \,\big\}&=2i\vd_\pp,  \label{eSusyD}\\
  \vd_{--}\id\vd_\mm&\Defl(\vd_\t{-}\vd_\s), &\qquad
  \vd_{++}\id\vd_\pp&\Defl(\vd_\t{+}\vd_\s),
\end{alignat}
and where $H=i\hbar\vd_\t$ and $p=-i\hbar\vd_\s$ are the worldsheet Hamiltonian and linear momentum, respectively. All other (anti)commutators among these operators vanish. In this notation, all operators\eq{e:SuSy} are eigen-operators of the Lorentz symmetry and the number of ``$\pm$'' sub/superscripts counts the additive eigenvalue in units of $\inv2\hbar$. So, the Lorentz-eigenvalue (``spin'') of $Q_\pm$ and $\Qb_\pm$ is $\pm\inv2\hbar$, and of $\vd_\pp$ is $+\hbar$, and of $\vd_\mm$ is $-\hbar$; also, $X^\pm\simeq X_\mp$, for any $X$.
\end{subequations}

Superfields defined solely by a pair of simple first-order superdifferential constraints, such as:
\begin{subequations}
 \label{e:HSF}
\begin{alignat}{5}
\text{chiral}&:&\qquad
  \Db_-\BF&=0,&\qquad \Db_+\BF&=0, \label{e:C}\\
\text{lefton}&:&\qquad
    D_-\BL&=0,&\qquad \Db_-\BL&=0, \label{e:L}\\
\text{righton}&:&\qquad
    D_+\bY&=0,&\qquad \Db_+\bY&=0, \label{e:R}
\end{alignat}
\end{subequations}
are some of the {\itshape haploid\/} superfields\cite{rHSS}. Unlike the well-known chiral super-constraints~\eq{e:C} and their lesser-known twisted kin\cite{rTwSJG0}, the defining super-constraints\eqs{e:L}{e:R} also {\itshape contain\/}\cite{rHSS} the worldsheet differential conditions:
\begin{equation}
 \big\{\,D_-\BL=0=\Db_-\BL\,\big\}~\supset~\vd_\mm\BL=0,\quad\hbox{and}\quad
 \big\{\,D_+\bY=0=\Db_+\bY\,\big\}~\supset~\vd_\pp\bY=0. \label{e:LR}
\end{equation}
These are then {\itshape partly\/} on-shell ({\em\/on half-shell\/}) representations of $(1,1|2,2)$-supersymmetry. Following the nomenclature of Ref.\cite{rUDSS01}, Ref.\cite{rHSS} jointly dubbed them {\itshape unidexterous\/}: $\BL$ a {\itshape lefton\/}, and $\bY$ a {\itshape righton\/}, since the component fields of $\BL=\BL(\s^\pp)$ move only to the left, whereas those of $\bY=\bY(\s^\mm)$ move only to the right.
 The constraints~\eqs{e:L}{e:R} allow for both complex and real $\BL$ and $\bY$; for simplicity, we will herein assume them to be real.

Berezin superintegrals are by definition equivalent to partial superderivatives, and up to total worldsheet derivatives (the worldsheet integrals of which are assumed to vanish, as usual) equivalent to covariant superderivatives~\cite{r1001,rPW,rWB,rBK}. Following Ref.\cite{rHSS}, we use (see appendix~A for further details):
\begin{equation}
 \int\rd^4\vs~(\3) ~:=~
 \inv8\Big(\big\{[\Db_+,D_+],[\Db_-,D_-]\big\}\3\Big)\Big| ~=:~(D^4\3)\big|,
 \label{e:DI}
\end{equation}
where the trailing ``$|$'' denotes projecting on the non-nilpotent part of superspace.
 Supersymmetric Lagrangians are then constructed following the prescription:
\begin{trivlist}\item[]\narrower\sl
 Let $\cBL$ be a functional expression of given superfields and their superderivatives. Let $\{D_1,\cdots,D_k\}$ be a basis of linear combinations of $D_\pm,\Db_\pm$, which do not annihilate $\int\rd^2\s\>\cBL$. Then
 \begin{equation}
 \cA:=\int\rd^2\s~\SL,
 \qquad\text{with}\qquad
 \SL~:=~ \big(\,D_{[1}\cdots D_{k]}~\cBL\,\big)\big|,
 \label{e:cL}
\end{equation}
is supersymmetric: $e^{-i(\e{\cdot}Q+\bar\e{\cdot}\Qb)}(\cA)=\cA$.
 $D_{[1}\cdots D_{k]}$ is the weighted antisymmetrized product.
\end{trivlist}
This is a formal and general statement of the well-known recipe\cite{r1001,rBK}. Its case-by-case proofs are scattered throughout the literature; a general and complete proof including cases with all types of gauge symmetries may be found in\cite{rIPd}.

Finally, we define the components of the superfields of interest as:
\begin{align}
  \El&:=\BL\big|,
 &\lb_+&:=D_+\BL\big|,
 &\bl_+&:=\Db_+\BL\big|,
 &\Lb_\pp&:=\inv2[\Db_+,D_+]\BL\big|, \label{e:Lc}\\
 \f&:=\BF\big|,
 &\j_+&:=D_+\BF\big|,
 &\j_-&:=D_-\BF\big|,
 &F&:=D_+D_-\BF\big|, \label{e:Rc}
\end{align}
and so on. For the most part, we will study $\BL$; analogous results follow for $\bY$.

\section{Lefton Superfields and Constrained Field-Space}
 \label{s:MO}
Consider a generic worldsheet lagrangian of the form
\begin{equation}
 \SL_\BL=\inv4\big[[\Db_+,D_+][\Db_-,D_-]\,\BK(\BL;\bX)\big]\big|,
  \label{e:LL}
\end{equation}
with $\BK(\BL;\bX)$ a suitable scalar functional expression involving $N$ lefton superfields $\BL^a$, and a total of $M$ other superfields, which we collectively denote by $\bX^i$.
 In the $(1,1)$-dimensional worldsheet, scalar fields the equations of motion of which linearize to the usual Klein-Gordon equation must have mass-dimension 0. As the lowest components of the superfields\eq{e:HSF} are all meant to be such propagating scalars, set $[\BF]=[\BL]=[\bs\Y]=0$.
 In turn, the definition of the action\eq{e:cL} with\eq{e:LL} implies that for $[\mathscr{A}]=0$, we need $[\BK]=0$, and so
\begin{equation}
 \left[\,\pd{\BK~~}{(D_+\BL^a)}\,\right]=-\inv2=\left[\,\pd{\BK~~}{(\Db_+\BL^a)}\,\right]
 \qquad\text{and}\qquad
 \left[\,\pd{\BK~~}{(\text{\small[}\Db_+,D_+\text{\small]}\BL^a)}\,\right]=-1.
\end{equation}
To avoid coefficients in the Lagrangian with negative mass-dimension (and an ensuing $M^{-n}$-type suppression by a mass-scale $M$), we must assume that $\BK(\BL;\bX)$ may depend on $\BL^a,\bX^i$, but not on their superderivatives.
 
A straightforward expansion of~\eq{e:LL} then yields:
\begin{equation}
 \SL_\BL = \Lb^a_\pp\,k^\pp_a+\inv2\bl^a_+\lb^b_+\,k^\pp_{ab}
           +\bl^a_+\,\k^+_a-\lb^a_+\,\vk^+_a+\sK,
 \label{e:LagL}
\end{equation}
as derived in \Eq{e:D4K}.
 The coefficients $k^\pp_a,k^\pp_{ab},\k^+_a,\vk^+_a$ and $\sK$ may well be functions of $\El^a=\BL^a|$, but not of $\Lb^a_\pp,\bl^a_+,\lb^a_+$: 
\begin{subequations}
 \label{e:Coefs}
\begin{alignat}{5}
 k^\pp_a
  &=\Big(\inv2[\Db_-,D_-]\,\pd{\BK~}{\BL^a}\Big)\Big|,&&&\qquad
 \k^+_a
  &=\Big(\inv2\Dh_+[\Db_-,D_-]\,\pd{\BK~}{\BL^a}\Big)\Big|,\\[2mm]
 k^\pp_{ab}
  &=\Big(\inv2[\Db_-,D_-]\,\ppd{\BK~}{\BL^a}{\BL^b}\Big)\Big|,&&&\qquad
 \vk^+_a
  &=\Big(\inv2\Dbh_+[\Db_-,D_-]\,\pd{\BK~}{\BL^a}\Big)\Big|,\\[2mm]
 \sK
  &=\inv4[\Dbh_+,\Dh_+][\Db_-,D_-]\,\BK(\BL;\bX)\big|,&&&\qquad
  \Dh_+\BL^a&=0=\Dbh_+\BL^a.
\end{alignat}
and let, subsequently,
\begin{equation}
 k^\pp_{b,a}\Defl\pd{k^\pp_b}{\El^a},\qquad
 k^\pp_{bc,a}\Defl\pd{k^\pp_{bc}}{\El^a},\qquad
  \k^+_{b,a}\Defl\pd{\k^+_b}{\El^a},\qquad
 \vk^+_{b,a}\Defl\pd{\vk^+_b}{\El^a},\qquad
 \sK_{\[a}\Defl\pd{\sK}{\El^a}.
\end{equation}
\end{subequations}
 It follows that the equations of motion for the component fields of $\BL^a$ are:
\begin{subequations}
 \label{e:LEoM}
\begin{alignat}{3}
 \d \Lb^a_\pp&:&\qquad
 k^\pp_a&=0, \label{e:k++}\\
 \d \bl^a_+&:&\qquad
 \inv2\lb^b_+\,k^\pp_{ab}+\k^+_a&=0, \label{e:k+}\\
 \d \lb^a_+&:&\qquad
 \inv2\bl^b_+\,k^\pp_{ab}+\vk^+_a&=0, \label{e:vk+}\\
 \d \El^a&:&\qquad
 \Lb^b_\pp\,k^\pp_{b,a}+\inv2\bl^b_+\lb^c_+\,k^\pp_{bc,a}
           +\bl^b_+\,\k^+_{b,a}-\lb^b_+\,\vk^+_{b,a}+\sK_{\[a}&=0. \label{e:sK}
\end{alignat}
\end{subequations}

If $\BK(\BL,\bX)$ is at least quadratic in the $\BL^a$, the functions\eq{e:Coefs} will also depend on the $\El^a$ but still not on the derivatives of $\El^a$. Also, the functions\eq{e:Coefs} all depend on the components of $\bX^i$ and the derivatives of some of them: On dimensional grounds and assuming that $[\bX^i]=0$, integration by parts may be used to ensure that the $\vd_\mm,\vd_\pp$-derivatives of only $x^i\Defl\bX^i|$, $\c^i_\pm\Defl D_\pm\bX^i|$ and $\x^i_\pm\Defl\Db_\pm\bX^i|$ occur in\eq{e:LagL} and the constraints\eq{e:LEoM}. Thus, all the other component fields of $\bX^i$ also have non-differential equations of motion. Component fields the equations of motion of which are not differential but algebraic\ft{\label{fn:Algebraic}In the $(1,1)$-dimensional worldsheet, a Lagrangian may well depend {\em\/transcendentally\/} on scalar fields. To avoid complicated circumlocutions, we imply any and all non-derivative dependence in ``algebraic equation of motion''.} are called {\em\/auxiliary\/}\cite{r1001,rPW,rWB,rBK}.

Thus, the system\eq{e:LEoM}|duly augmented by the $M'$ (possibly complex) equations of motion of the total of $M'$ (possibly complex) auxiliary component fields in $\bX^i$| may be regarded as a system of {\em\/algebraic\/} equations of motion over the $4N$-dimensional field space $\big\{\El^a,\lb^a_+,\bl^a_+,\Lb^a_\pp;\dots\big\}$, where the ellipses stand for the auxiliary fields of the $\bX^i$. We have that\ft{\label{fn:Bigger}This assumes that superfields are used to represent {\em\/simple\/} representations of supersymmetry, obtained by constraining a complex, otherwise unconstrained and unprojected superfield. It {\em\/is\/} possible to define indefinitely larger and more complex representations\cite{r6-1,r6-4} all of which {\em\/can\/} be realized in terms of superfields following Ref.\cite{r6-1.2}, requiring however a possibly indefinite array of simple superfields.}
\begin{equation}
  0 \leq M' \leq 16M,
\end{equation}
since the smallest nontrivial superfields has a single, propagating boson-fermion pair and no auxiliary component field|the minimal number, and $(1,1|2,2)$-supersymmetry affords at most sixteen component fields per {\em\/simple\/} superfield$^\textsl{\ref{fn:Bigger}}$\cite{rHSS}.

The space of simultaneous solutions of the equations of motion\eq{e:LEoM}|together with the $M'$ (possibly complex) equations of motion of the total of $M'$ (possibly complex) auxiliary component fields in $\bX^i$|is then an {\em\/essentially algebraic\/}$^\textsl{\ref{fn:Algebraic}}$ family of varieties.

The generic member of this family is parametrized by the dynamical scalars $x^i$, with the dynamical of the fermions $\c^i_\pm,\x^i_\pm$ spanning copies of the (co)tangent bundles. With this in mind, it is standard\cite{r1001,rPW,rWB,rPhases,rBK} to replace the auxiliary fields with the solutions of their equations of motion and so simplify the Lagrangian. The main point of the present analysis is that the space of solutions to these equations is {\em\/straitified\/}, and may also contain both conjoined and disjoint components. It would seem to us that such families of target spaces naturally incorporate the so-called brane-Worlds that have received considerable interest in the past decade. That such stratified families of target spaces arise {\em\/naturally\/} in worldsheet field theories has been noted over a decade ago\cite{rHitch}.

In particular, the essentially algebraic family of varieties obtained as the solution of the auxiliary fields' equations of motion may well contain sub-generic members, which are radically different from the generic model. It is the purpose of this note to illustrate the emergence of this nontrivial topological and geometric structure in target spaces by a few explicit examples that follow. A more comprehensive analysis is found in Ref.\cite{rIPd}.

\section{A Simple Example}
 \label{s:SX}
Rather than reproducing the complete analysis\cite{rIPd}, consider a system of $M$ chiral superfields $\BF^i$ ($\BFb_i\Defl\6(\BF^i){}^\dag$) and a single lefton superfield, $\BL$, equipped with a Lagrangian of the form
\begin{align}
 &\SL_{N,1}
  \Defl\inv4[\Db_+,D_+][\Db_-,D_-]
    \big(\d^i_j+\BL\,h^i{}_j\big)(\BFb_i\,\BF^j)\big|,\nn\\
 &=\Lb_\pp\,h^i{}_j\,\big[\jb_{-i}\j^j_-
       +i(\fb_i\dvd_\mm\f^j)\big]
  -\bl_+\,h^i{}_j\,\big[\jb_{-i}F^j
       -i(\fb_i\dvd_\mm\j^j_+)\big]
  +\lb_+\,h^i{}_j\,\big[\Fb_i\j^j_-
       -i(\jb_{+i}\dvd_\mm\f^j)\big]\nn\\
  &\quad
  +\big[\d^i_j+\El\,h^i{}_j\big]
     \big[\Fb_i F^j
          +i(\jb_{-i}\dvd_\pp \j^j_-)
          +i(\jb_{+i}\dvd_\mm \j^j_+)
          +2(\vd_\mm\fb_i)(\vd_\pp\f^j)
          +2(\vd_\pp\fb_i)(\vd_\mm\f^j)\big],
  \label{e:L1}
\end{align}
where we have dropped total derivative terms, including
\begin{equation}
 \El\,h^i{}_j\,\vd_\mm\vd_\pp(\fb_i\f^j)
 =\vd_\mm\big(\El\,h^i{}_j\,\vd_\pp(\fb_i\f^j)\big)
  -h^i{}_j\,(\underbrace{\vd_\mm\El}_{=0})\,\vd_\pp(\fb_i\f^j).
\end{equation}

Equations of motion may be obtained from the general formula:
\begin{equation}
 \d X:\quad
  0=\pd{\SL}{X}-\vd_\pp\pd{\SL}{(\vd_\pp X)}-\vd_\mm\pd{\SL}{(\vd_\mm X)},
\end{equation}
and they are particularly simple for the components of $\BL$:
\begin{subequations}
 \label{e:L1EoM}
\begin{alignat}{3}
 \d\Lb_\pp&:&\quad
0&=h^i{}_j\,\big[\jb_{-i}\j^j_-
       +i(\fb_i\dvd_\mm\f^j)\big], \label{e:L1EoM2}\\
 \d\bl_+&:&\quad
0&=h^i{}_j\,\big[\jb_{-i}F^j
       -i(\fb_i\dvd_\mm\j^j_+)\big], \label{e:L1EoM1b}\\
 \d\lb_+&:&\quad
0&=h^i{}_j\,\big[\Fb_i\j^j_-
       -i(\jb_{+i}\dvd_\mm\f^j)\big], \label{e:L1EoM1}\\
 \d\El&:&\quad
0&=h^i{}_j
     \big[\Fb_i F^j+\cK_i{}^j\big], \label{e:L1EoM0}
\end{alignat}
where the trace of
\begin{equation}
 \cK_i{}^j\Defl
           i\big[(\jb_{-i}\dvd_\pp \j^j_-)
                +(\jb_{+i}\dvd_\mm \j^j_+)\big]
          +2\big[(\vd_\mm\fb_i)(\vd_\pp\f^j)
                +(\vd_\pp\fb_i)(\vd_\mm\f^j)\big]
\end{equation}
\end{subequations}
is the ``standard kinetic term''.
Upon enforcing these constraints (by integrating the partition functional over $\BL$), the components of $\BL$ disappear from the Lagrangian, which then has the form:
\begin{equation}
 \SL_{N,1}\big|_{\text{(\ref{e:L1EoM})}}
 =\big[\Fb_i F^i
          +i(\jb_{-i}\dvd_\pp \j^i_-)
          +i(\jb_{+i}\dvd_\mm \j^i_+)
          +2(\vd_\mm\fb_i)(\vd_\pp\f^i)
          +2(\vd_\pp\fb_i)(\vd_\mm\f^i)\big]\Big|_{\text{(\ref{e:L1EoM})}},
 \label{e:L10}
\end{equation}
where $\f^i,\j^i_\pm,F^i$ and their conjugates are subject to the constraints\eq{e:L1EoM}, and which cannot|for general $h^i{}_j$|be solved straightforwardly to simplify the Lagrangian\eq{e:L10} any further. Owing to these enforced constraints, the simple appearance of, say, $\Fb_iF^i$ is misleading: the equations of motion for $\Fb_i$ and $F^j$ are not simply $\Fb_i=0=F^i$ as the form of\eq{e:L10} would seem to suggest. Instead, one must pursue the considerably more involved {\em\/constrained\/} variations, subject to\eq{e:L1EoM}.

\subsection{A Single Chiral Superfield}
 \label{s:1}
As an illustration, consider the ``near-trivial'' case, with a single $(\BF,\BFb)$ pair, so $M=1$. Eqs.\eq{e:L1EoM} simplify and are amended by:
\begin{subequations}
\begin{alignat}{3}
 \d\Fb&:&\quad
 h\,\lb_+\,\j_-  &= -(1+h\,\El)F, \label{e:L1EoMFb}\\
 \d F&:&\quad
 h\,\bl_+\,\jb_-  &= +(1+h\,\El)\Fb. \label{e:L1EoMF}
\end{alignat}
\end{subequations}
Assuming that $\El\neq-1/h$ and using the latter two equations (\ie, integrating out $F,\Fb$ first) enforces
\begin{equation}
 F   =-\frac{\lb_+\,\j_-}{1+h\,\El}\,h,\qquad\text{and}\qquad
 \Fb =+\frac{\bl_+\,\jb_-}{1+h\,\El}\,h.
 \label{e:F+Fb}
\end{equation}
Whereas dividing by a field is ill-defined in {\em\/quantum\/} field theory in general, $(1+h\,\El)^{-1}$ may be understood in terms of an infinite geometric series. Substituting\eq{e:F+Fb} into\eq{e:L1} produces
\begin{align}
 \SL_{1,1}\big|_{F,\Fb}
 &=h\,\Lb_\pp\big[\jb_-\j_-
       +i(\fb\dvd_\mm\f)\big]
  +ih\,\bl_+(\fb\dvd_\mm\j_+)
  -ih\,\lb_+(\jb_+\dvd_\mm\f)
  -\frac{\bl_+\lb_+\jb_-\j_-}{1+h\,\El}\,h^2\nn\\*
  &\quad
  +(1+h\,\El)
     \big[i(\jb_-\dvd_\pp \j_-)
          +i(\jb_+\dvd_\mm \j_+)
          +2(\vd_\mm\fb)(\vd_\pp\f)
          +2(\vd_\pp\fb)(\vd_\mm\f)\big],
  \label{e:L1FF}
\end{align}
where $(1+h\,\El)^{-1}$ should again be understood as abbreviating the infinite geometric series.
Variation over the components of $\BL$ now enforces the constraints:
\begin{subequations}
\begin{alignat}{3}
 \d\Lb_\pp&:&\quad
 \jb_-\j_-
 &=-i(\fb\dvd_\mm\f), \label{e:L1EoM4}\\
 \d\bl_+&:&\quad
 h\,\lb_+\jb_-\j_-
 &=i(1+h\,\El)(\fb\dvd_\mm\j_+), \label{e:FFb}\\
 \d\lb_+&:&\quad
 h\,\bl_+\jb_-\j_-
 &=i(1+h\,\El)(\jb_+\dvd_\mm\f), \label{e:FFa}\\
 \d\El&:&\quad
 h^2\bl_+\lb_+\jb_-\j_-
 &=-(1+h\,\El)^2\,\cK,\label{e:FF0}\\
 &&\cK
 &\Defl
     \big[i(\jb_-\dvd_\pp \j_-)
          +i(\jb_+\dvd_\mm \j_+)
          +2(\vd_\mm\fb)(\vd_\pp\f)
          +2(\vd_\pp\fb)(\vd_\mm\f)\big],
\end{alignat}
\end{subequations}
where the right-hand side of each equation is dynamical.

Note that \Eq{e:L1EoM4} is a dynamical constraint purely on the $\sF=(\f,\fb;\j_\pm,\jb_\pm;F,\Fb)$ field space spanned by the components of $\BF,\BFb$. In turn, the coupled system\eqs{e:FFb}{e:FF0} formally determines $\lb_+,\bl_+,\El$ in terms of the components of $\BFb,\BF$, and $\Lb_\pp$ remains unconstrained and uncoupled.

\paragraph{Wherever $\jb_-\j_-\neq0$:}
Taking the ratio of\eq{e:FFa} by\eq{e:FFb} produces the formal solution
\begin{equation}
 \bl_+=\lb_+\bigg(\frac{\jb_+\dvd_\mm\f}{\strut\fb\dvd_\mm\j_+}\bigg),
 \label{e:ll}
\end{equation}
the substitution of which in\eq{e:L1FF} produces
\begin{align}
 \SL_{1,1}\big|_{F,\Fb,\bl_+,\lb_+}
 &=h\,\Lb_\pp\big[\jb_-\j_-
       +i(\fb\dvd_\mm\f)\big]
  +(1+h\,\El)\,\cK,
\end{align}
where we have also used that $\bl_+\lb_+\propto(\lb_+)^2\id0$. Finally, integrating $\Lb_\pp$ out continues to impose the constraint\eq{e:L1EoM4}, while integrating $\El$ out imposes the constraint $\text{Tr}[\cK]=0$, thus forcing the standard kinetic term to vanish. This reduces the entire $\sF$ field space to non-dynamical constants.

\paragraph{Wherever $\jb_-\j_-=0$:}
The left-hand side of each of the constraints\eqs{e:L1EoM4}{e:FF0} vanishes. In the resulting constraints, $\El\neq-1/h$ would again imply $\text{Tr}[\cK]=0$, which reduces the entire field space to non-dynamical constants.

\begin{figure}[ht]
 \begin{center}
  \begin{picture}(120,72)(0,8)
   \put(0,0){\includegraphics[width=120mm]{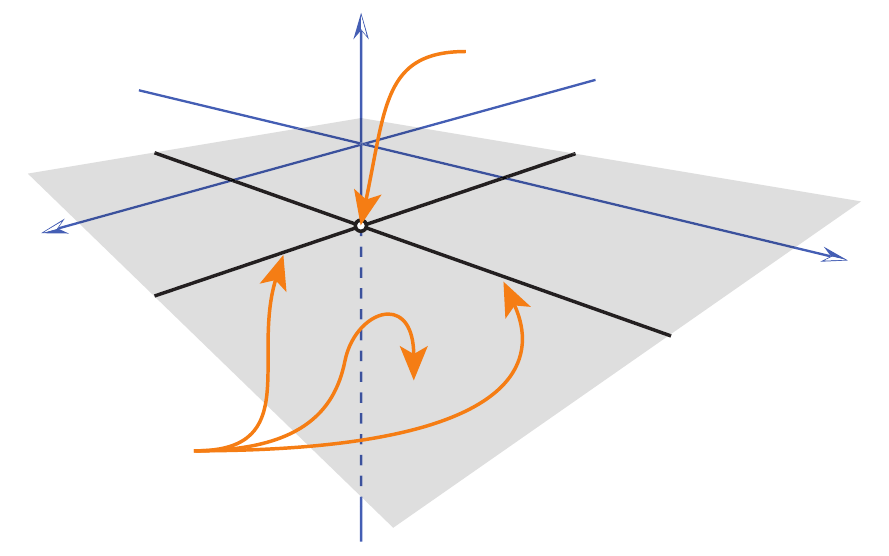}}
   \put(-10,70){\parbox{50mm}{\raggedright
        The $(\El;\lb_+,\bl_+)\subset\BL$ subset of the $\BL$-field space.}}
   \put(3,13){Eqs.\eq{e:11}}
   \put(64,67){Eqs.\eq{e:00}}
   \put(84,50){The $\El=-1/h$ plane}
   \put(5,40){$\lb_+$}
   \put(109,37){$\bl_+$}
   \put(45,70){$\El$}
  \end{picture}
 \end{center}
 \caption{A sketch of the situation in the $\BL=(\El;\lb_+,\bl_+;\Lb_\pp)$ field space described by \Eqs{e:L10}{e:11}: The $\sF=(\f,\fb;\jb_\pm,\j_\pm;F,\Fb)$ field space is constrained to be constant away from the indicated plane $\El=-1/h$, is least constrained when $\lb_+=0=\bl_+$, and satisfies the intermediate constraint system\eq{e:00} in the rest of the $\El=-1/h$ plane, with the $\lb_+$- and $\bl_+$-axes included. The constrained $\sF$ field space is thus fibered over the $\BL$ field space: $\sF$ is essentially trivial away from the $\El=-1/h$ plane, and subject to\eq{e:00} and\eq{e:11} otherwise.}
 \label{f:L1FS}
\end{figure}

\paragraph{Wherever $\El=-1/h$:}
On the special field-space locus $\El=-1/h$, the component field $\El$ decouples, \Eq{e:L1EoM0} becomes void, and now \Eqs{e:L1EoM2}{e:L1EoM1} and\eqs{e:L1EoMFb}{e:L1EoMF} reduce to:
\begin{subequations}
\begin{align}
 \jb_-\j_-       &=-i(\fb\,\dvd_\mm\f),\\
 \jb_-F          &=i(\fb\,\dvd_\mm\j_+),\\
 \Fb\j_-         &=i(\jb_+\,\dvd_\mm\f),\\
 \lb_+\j_-   &=0,\\
 \bl_+\jb_- &=0.
\end{align}
\end{subequations}
Depending on the vanishing of $\bl_+,\lb_+$, we have the cases:
\begin{subequations}
\begin{description}\itemsep=-3pt\vspace{-2mm}
 \item[\hglue2pc{$\bullet~(\bl_+,\lb_+)=(0,0)$}]
  \begin{equation}
    \jb_-\j_-=-i(\fb\,\dvd_\mm\f),\qquad
    \jb_-F=i(\fb\,\dvd_\mm\j_+),\qquad
    \Fb\j_-=i(\jb_+\,\dvd_\mm\f).
    \label{e:00}
  \end{equation}
 \item[\hglue2pc{$\bullet~\bl_+=0\neq\lb_+$\,:}]
  \begin{equation}
    \j_-=0,\qquad
    (\fb\,\dvd_\mm\f)=0,\qquad
    \jb_-F=i(\fb\,\dvd_\mm\j_+),\qquad
    (\jb_+\,\dvd_\mm\f)=0.
    \label{e:01}
  \end{equation}
 \item[\hglue2pc{$\bullet~\bl_+\neq0=\lb_+$\,:}]
  \begin{equation}
    \jb_-=0,\qquad
    (\fb\,\dvd_\mm\f)=0,\qquad
    (\fb\,\dvd_\mm\j_+)=0,\qquad
    \Fb\j_-=i(\jb_+\,\dvd_\mm\f).
    \label{e:10}
  \end{equation}
Since $\BFb=\BF^\dag$, $(\j_-{=}0)\Iff(\jb_-{=}0)$, so that the constraint systems\eqs{e:01}{e:10} in $\sF$ both coalesce to the result obtained when $\lb_+\neq0\neq\bl_+$. We thus only have one other case:
 \item[\hglue2pc{$\bullet~(\bl_+,\lb_+)\neq(0,0)$\,:}]
  \begin{equation}
    \jb_-,\j_-=0,\qquad
    (\fb\,\dvd_\mm\f)=0,\qquad
    (\fb\,\dvd_\mm\j_+)=0,\qquad
    (\jb_+\,\dvd_\mm\f)=0.
    \label{e:11}
  \end{equation}
\end{description}
\end{subequations}
The location in the $\BL$-space of the resulting {\em\/dynamical\/} constraint system\eq{e:00} and\eq{e:11} on the $\sF$ field space is sketched in Fig.~\ref{f:L1FS}.

\subsection{A Physics Interpretation}
By writing $\f=\f_1+i\f_2$, we obtain:
\begin{equation}
 -i(\fb\,\dvd_\mm\f) = 2(\f_1\dvd_\mm\f_2)
   = \2{(\f_1\dot\f_2-\dot\f_1\f_2)}-(\f_1\acute\f_2-\acute\f_1\f_2),
 \label{e:LH}
\end{equation}
where over-dots denote derivatives with respect to time as usual, and over-primes are derivatives with respect to space. The underlined term is recognized to be the the conventional angular momentum in the $(\f_1,\f_2)$-plane.
 Thus, $-i(\fb\,\dvd_\mm\f)$ generalizes the angular momentum in the complex $\f$-plane from (quantum) mechanics on the worldline.

To be precise, \Eq{e:LH} provides a definition for the ``right-handed'' generalization of angular momentum, whereas the analogous expression with $\dvd_\pp$ instead of $\dvd_\mm$, and so a positive relative sign between the last two terms in\eq{e:LH}, would be a ``left-handed'' generalization. In turn, it is also possible to identify the worldsheet light-cone coordinate $\s^\mm\Defl\t{-}\s$ as a ``time'', with respect to which $-i(\fb\,\dvd_\mm\f)$ is the unadulterated angular momentum in the $(\f_1,\f_2)$-plane.

It is then clear that the Lagrangian\eq{e:L1} couples the $\Lb_\pp$ component field of the lefton superfield $\BL$ to the supersymmetric completion of the right-handed generalization of the angular momentum, and the remaining $h$-dependent terms in that Lagrangian simply provide for the $(1,1|2,2)$-supersymmetric completion of this coupling. This is curiously similar to the couplings examined in Ref.\cite{r6-7a}, wherein the supersymmetric completion\ft{This is with respect to the $N=2$-extended worldline supersymmetry.} of the conventional angular momentum, $\big[(\f_1\dvd_\t\f_2)+2i\j_1\j_2\big]$, was coupled to {\em\/external\/} magnetic fluxes.

Comparing our Lagrangians with those in Ref.\cite{r6-7a}, we see that the component field $\Lb_\pp$ herein stands in for the external magnetic flux of Ref.\cite{r6-7a}. The toy model considered above thus:
\begin{enumerate}\itemsep=-3pt\vspace{-2mm}
 \item provides a $(1,1|2,2)$-supersymmetric generalization of the Zeeman couplings from Ref.\cite{r6-7a};
 \item identifies the external magnetic flux of Ref.\cite{r6-7a} with the worldsheet component field $\Lb_\pp\subset\BL$, or perhaps more properly, with its vacuum-expectation value;
 \item assigns $\El,\lb_+,\bl_+\subset\BL$ as the super-partners of this ``imported'' magnetic flux;
 \item subjects these ``imported'' super-flux variables, $\El,\bl_+,\lb_+,\Lb_\pp$, also to the least action principle.
\end{enumerate}
Consistently with standing in for an {\em\/external\/} magnetic flux of Ref.\cite{r6-7a}, the component field $\Lb_\pp$ herein remains a non-propagating degree of freedom, as are its super-partners, $\El,\lb_+$ and $\bl_+$.

\begin{figure}[ht]
 \begin{center}
  \begin{picture}(120,72)(0,8)
   \put(0,0){\includegraphics[width=120mm]{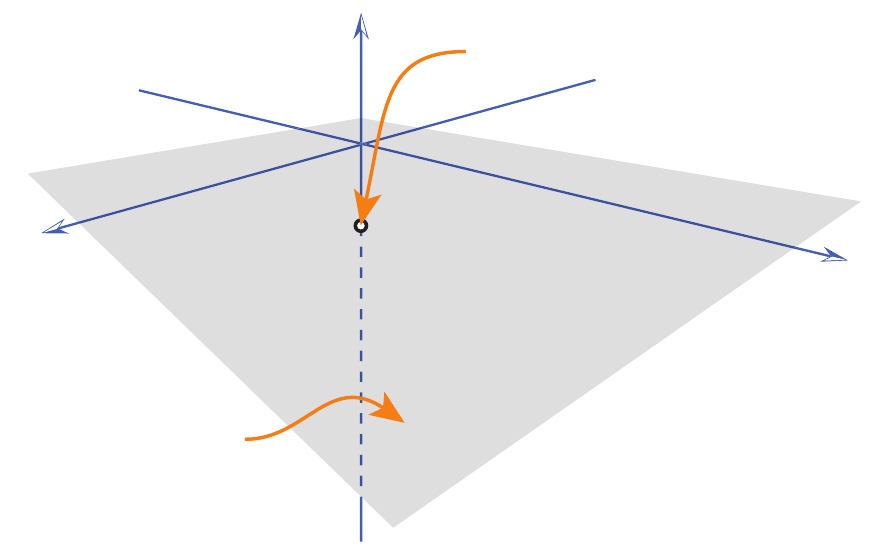}}
   \put(-10,70){\parbox{50mm}{\raggedright
        The $(\El;\lb_+,\bl_+)\subset\BL$ subset of the $\BL$-field space.}}
   \put(17,15){Phase \bf B}
   \put(6,10){(The $\El=-1/h$ plane)}
   \put(64,67){Phase \bf C}
   \put(87,15){Phase \bf A}
   \put(70,10){(Outside the $\El=-1/h$ plane)}
   \put(5,40){$\lb_+$}
   \put(109,37){$\bl_+$}
   \put(45,70){$\El$}
  \end{picture}
 \end{center}
 \caption{A sketch of the situation in the phase diagram of an $N{=}1$ model\eq{e:L1}.}
 \label{f:L1PhD}
\end{figure}

Owing to the addition (in comparison with the work of Ref.\cite{r6-7a}) of $\El,\lb_+$ and $\bl_+$, which are super-partners of $\Lb_\pp$ with respect to the worldsheet $(2,2)$-supersymmetry, we now have the nontrivial field-space sketched in Fig.~\ref{f:L1FS}, with three physically distinct ``phases'' stitched together:
\begin{description}\itemsep=-3pt\vspace{-2mm}
 \item[\hglue1.5pc\bf A.][$\El\neq-1/h$]:
   The standard kinetic terms are forced to vanish, so that there is no dynamics.
 \item[\hglue1.5pc\bf B.][$\El=-1/h$, $(\lb_+,\bl_+)\neq(0,0)$]:
   The dynamical conditions\eq{e:11} restrict the field-space $\sF=(\f,\fb;\j_\pm,\jb_\pm;F,\Fb)$ in a left-right asymmetric fashion: $\j_-=0=\jb_-$, the component fields $\f,\fb,\j_+,\jb_+$ must satisfy
\begin{equation}
   (\fb\,\dvd_\mm\f)=(\fb\,\dvd_\mm\j_+)=(\jb_+\,\dvd_\mm\f)=0,
\end{equation}
and $F$ and $\Fb$ decouple.
 \item[\hglue1.5pc\bf C.][$\El=-1/h$, $(\lb_+,\bl_+)=(0,0)$]:
   The dynamical conditions\eq{e:00} restrict the field-space $\sF=(\f,\fb;\j_\pm,\jb_\pm;F,\Fb)$ in a left-right asymmetric fashion:
\begin{equation}
   \jb_-\j_-=-i(\fb\,\dvd_\mm\f),\qquad
    \jb_-F=i(\fb\,\dvd_\mm\j_+),\qquad
    \Fb\j_-=i(\jb_+\,\dvd_\mm\f).
\end{equation}
\end{description}
The complete model governed by the Lagrangian\eq{e:L1} with a single pair of chiral superfield, $\BF,\BFb$, then {\em\/contains\/} all three phases, fibered over the $\BL$-field space, and for which the $(\El;\lb_+,\bl_+)$-space  sketch in Fig.~\ref{f:L1FS} becomes a {\em\/de facto\/} phase diagram, shown in Fig.~\ref{f:L1PhD}.
This hierarchical structure, {\em\/stratification\/}, of the field space reminds of brane-world scenarios, in that:
\begin{enumerate}\itemsep=-3pt\vspace{-2mm}
 \item the dimension of the field space region (stratum) corresponding to a phase may well vary from phase to phase: the various strata may well have various dimensions;
 \item not just the dimension (number of dynamical degrees of freedom), but also the dynamics itself may well vary from phase to phase, \ie, from stratum to stratum.
\end{enumerate}

\subsection{More Superfields}
The foregoing analysis demonstrates the emergence of a non-trivial constrained field space already in the simple case involving a single chiral superfield, its Hermitian conjugate, and a single real lefton superfield, which furthermore was restricted to occur linearly in the Lagrangian. Evidently, much more complicated geometries may be achieved by increasing the number and diversity of superfields involved, and including more non-linear expressions in the Lagrangians.

Revisiting the Lagrangian\eq{e:L1}, setting now $N=2$ and selecting the matrix $h^i{}_j$ to be of the form $[h^i{}_j]=\sm{0&0\\0&h}$ embeds the above model in the field space of $(\BL;\BF^1,\BFb_1,\BF^2,\BFb_2)$: the pair $\BF^2,\BFb_2$ becomes constrained in the manner of the toy model of section~\ref{s:1}, whereas the pair $\BF^1,\BFb_1$ retains the usual free-field dynamics of a chiral superfield without a potential. Of course, the two sectors may easily be coupled through the introduction of a mixing terms in the superpotential $W(\BF^1,\BF^2)$ and/or by selecting an off-diagonal matrix $[h^i{}_j]$.

Pursuing the former option and considering a term (with $\El=-1/h$)
\begin{equation}
 [D_+,D_-]\,w_{12}\BF^1\BF^2|+\textit{h.c.},
 ~=~w_{12}\big\{F^1\f^2+\f^1F^2+\j_+^1\j_-^2+\j_+^2\j_-^1\big\}+\textit{h.c.},
 \label{e:Ww-}
\end{equation}
we notice that at $(\lb_+,\bl_+)=(0,0)$, this is indeed the contribution to the potential. However, for $(\lb_+,\bl_+)\neq(0,0)$ where \Eq{e:11}, with the above choice of $[h^i{}_j]$, implies that this contribution to the potential truncates to
\begin{equation}
 [D_+,D_-]\,w_{12}\BF^1\BF^2|+\textit{h.c.},
 ~=~w_{12}\big\{F^1\f^2+\f^1F^2+\j_+^2\j_-^1\big\}+\textit{h.c.}
 \label{e:Wwo-}
\end{equation}
The situation represented by the phase diagram in Fig.~\ref{f:L1FS} thus also affords a variation of interactions: the potential term\eq{e:Wwo-} changes to\eq{e:Ww-} in Phase~{\bf C}, where $(\lb_+,\lb_+)\to(0,0)$ and $\j^2_-=0=\jb_-^2$.

In the current set-up, this change is not dynamical, since the component fields of $\BL$ are non-propagating auxiliary fields. Nevertheless, the ``Universe'' described by any model containing Lagrangian terms like\eq{e:L1} and\eqs{e:Ww-}{e:Wwo-} will contain a hierarchically nested structure of phases, resembling the sketch in Fig.~\ref{f:L1PhD}.

\subsection{Nonlinearity}
Consider now a Lagrangian of the general form\eq{e:LL}, where $\BK(\BL;\bX)$ is now a quadratic function of a single lefton superfield $\BL$:
\begin{equation}
 \BK(\BL;\bX) = \BL\!^2\bA(\bX)+\BL\bB(\bX),
\end{equation}
with $\bA(\bX)$ and $\bB(\bX)$ suitable functions of the other superfields, $\bX$.
Then:
\begin{equation}
 \SL_\BL = \Lb_\pp(2\El\,A^\pp+B^\pp)
           +\bl_+\lb_+A^\pp
           +\bl_+(2\El\a^++\b^+)
           -\lb_+(2\El\bar\a^++\bar\b^+)
           +\El^2\sA+\El\sB,
 \label{e:LagL2}
\end{equation}
with the coefficients:
\begin{subequations}
 \label{e:CoeFs}
\begin{alignat}3
 A^\pp
  &= \inv2[\Db_-,D_-]\bA(\bX)\big|,&\qquad
 B^\pp
  &= \inv2[\Db_-,D_-]\bB(\bX)\big|,\\[2mm]
 \a^+
  &= \inv2D_+[\Db_-,D_-]\bA(\bX)\big|,&\qquad
 \b^+
  &= \inv2D_+[\Db_-,D_-]\bB(\bX)\big|,\\[2mm]
 \bar\a^+
  &= \inv2\Db_+[\Db_-,D_-]\bA(\bX)\big|,&\qquad
 \bar\b^+
  &= \inv2\Db_+[\Db_-,D_-]\bB(\bX)\big|,\\[2mm]
 \sA
  &=\inv4[\Db_+,D_+][\Db_-,D_-]\bA(\bX)\big|,&\qquad
 \sB
  &=\inv4[\Db_+,D_+][\Db_-,D_-]\bB(\bX)\big|.
\end{alignat}
\end{subequations}

The equations of motion for $(\El;\lb_+,\bl_+;\Lb_\pp)$ are:
\begin{subequations}
 \label{e:EoML2}
\begin{alignat}{3}
 \d\Lb_\pp&:&\qquad
 2\El\,A^\pp+B^\pp&=0,\\[2mm]
 \d\lb_+&:&\qquad
 \bl_+A^\pp+(2\El\bar\a^++\bar\b^+)&=0,\\[2mm]
 \d\bl_+&:&\qquad
 \lb_+A^\pp+(2\El\a^++\b^+)&=0,\\[2mm]
 \d\El&:&\qquad
 2\Lb_\pp A^\pp+2\bl_+\a^+-2\lb_+\bar\a^++2\El\sA+\sB&=0.
\end{alignat}

\end{subequations}
\paragraph{Wherever $A^\pp\neq0$:}
Solving them in turn implies that:
\begin{equation}
 \begin{gathered}
  \El   =-\frac{B^\pp}{2A^\pp},\qquad
 \lb_+  =\frac{B^\pp\a^+}{(A^\pp)^2}-\frac{\b^+}{A^\pp},\qquad
 \bl_+  =\frac{B^\pp\bar\a^+}{(A^\pp)^2}-\frac{\bar\b^+}{A^\pp},\\[2mm]
 \Lb_\pp=-2\frac{B^\pp\bar\a^+\a^+}{(A^\pp)^3}
          +\frac{\bar\b^+\a^++\bar\a^+\b^+}{(A^\pp)^2}
         +\frac{B^\pp\sA}{2(A^\pp)^2}
         -\frac{\sB}{2A^\pp},
 \end{gathered}
 \label{e:SolL2}
\end{equation}
Substituting these back into the Lagrangian\eq{e:LagL2} produces:
\begin{equation}
 \SL_\BL\big|
 =(A^\pp)^{-3}\big(B^\pp\a^+-A^\pp\b^+\big)\big(B^\pp\bar\a^+-A^\pp\bar\b^+\big)
 +\inv4(A^\pp)^{-2}B^\pp\big(B^\pp\sA-2A^\pp\sB\big),
 \label{e:NonStL}
\end{equation}
which is clearly non-standard. Consequently, the equations of motion derived from the Lagrangian\eq{e:NonStL} also contain non-standard terms owing to the factors $(A^\pp)^{-n}$.

In addition to the equations of motion derived from this Lagrangian, the fact that $\vd_\mm\BL=0$ implies that the solutions\eq{e:SolL2} themselves must be unidexterous:
\begin{equation}
 \begin{gathered}
  \vd_\mm\Big(\frac{B^\pp}{2A^\pp}\Big)=0,\qquad
  \vd_\mm\Big(\frac{B^\pp\a^+}{(A^\pp)^2}-\frac{\b^+}{A^\pp}\Big)=0,\qquad
  \vd_\mm\Big(\frac{B^\pp\bar\a^+}{(A^\pp)^2}-\frac{\bar\b^+}{A^\pp}\Big)=0,\\[2mm]
  \vd_\mm\Big(\frac{B^\pp\bar\a^+\a^+}{(A^\pp)^3}
          -\frac{\bar\b^+\a^++\bar\a^+\b^+}{(A^\pp)^2}
         -\frac{B^\pp\sA}{2(A^\pp)^2}
         +\frac{\sB}{2A^\pp}\Big)=0.
 \end{gathered}
 \label{e:UL2}
\end{equation}
Eqs.\eq{e:UL2} are then non-linear, {\em\/unidexterous\/} and {\em\/dynamical\/} constraints on the field space spanned by the components of $\bX$.

Both these additional, unidexterous constraints\eq{e:UL2} and the equations of motion derived from the non-standard Lagrangian\eq{e:NonStL} are all readily computed for any concrete choice of $\bA(\bX)$ and $\bB(\bX)$ using Eqs.\eq{e:CoeFs}.

\paragraph{Wherever $A^\pp=0$:}
In these regions of the field space $\sF$, Eqs.\eq{e:EoML2} simplify to:
\begin{equation}
 \begin{aligned}
   B^\pp&=0,&\qquad
   \bar\b^+&=-2\El\,\bar\a^+,\\
   \sB&=-2\bl_+\a^++2\lb_+\bar\a^+-2\El\sA,&\qquad
   \b^+&=-2\El\,\a^+,\\
 \end{aligned} 
 \label{e:Lin}
\end{equation}
without the coefficients $\a^+,\bar\a^+,\sA$ necessarily vanishing also.
 The first of these constraints, $B^\pp=0$ reduces $\sF$, whereas the remaining equations provide a linear fibration of $\sF|_{B^\pp=0}$ over $(\El;\lb_+,\bl_+)$: a structure that is in general quite different from that described by the Lagrangian\eq{e:NonStL} and the auxiliary left-right asymmetric dynamical constraints\eq{e:UL2}.

The Lagrangian\eq{e:LagL2} thus also specifies a stratified field space, with the special stratum this time being located within $\sF$, as the vanishing locus of $A^\pp$. Elsewhere, we have the left-right asymmetrically constrained\eq{e:UL2} dynamics governed by the Lagrangian\eq{e:NonStL}; at the vanishing locus of $A^\pp$, however, the field space acquires the geometry of the linear fibration\eq{e:Lin}.

A simple example of this (dynamical) stratification is given by choosing $\bA=\BFb\BF$, whereupon
\begin{subequations}
 \label{e:FFsol}
\begin{gather}
 A^\pp=\jb_-\j_-+i\big(\fb \dvd_\mm \f\big),\\
 \a^+=-\jb_-F+i\big(\fb \dvd_\mm \j_+\big),\qquad
 \bar\a^+=-\Fb\j_-+i\big(\jb_+ \dvd_\mm \f\big),\\
 \sA=\Fb F +i\big(\jb_-\dvd_\pp\j_-\big)+i\big(\jb_+\dvd_\mm\j_+\big)
     +2\big[(\vd_\mm\fb)(\vd_\pp\f)+(\vd_\pp\fb)(\vd_\mm\f)\big],
\end{gather}
\end{subequations}
and leaving $\bB$ arbitrary for now.
Substitution of\eq{e:FFsol} into\eq{e:NonStL} yields the non-standard Lagrangian dictating the dynamics away from the $\jb_-\j_-=-i\big(\fb \dvd_\mm \f\big)$ locus. This then is also amended by the left-right asymmetric conditions\eq{e:UL2}.

In turn, wherever in $\sF$ field space $\jb_-\j_-=-i\big(\fb \dvd_\mm \f\big)$, we have Eqs.\eq{e:Lin}, which constrain the components $B^\pp,\b^+,\bar\b^+$ and $\sB$ of the as yet unspecified super-function $\bB$, in terms of $\a^+,\bar\a^+,\sA$ and $\El$|and without any constraints of the left-right asymmetric form\eq{e:UL2}. This leads to a markedly different dynamics over this dynamically determined subregion of the field space $\sF$.

Recall that $A^\pp=\big[\jb_-\j_-+i\big(\fb \dvd_\mm \f\big)\big]$ is the supersymmetrized right-handed generalization of angular momentum in the $\f$-plane. This stratification is then, in physical terms, located at the vanishing locus of this supersymmetric physical observable.

\section{Conclusions}
 \label{s:C}
We have studied some aspects of coupling worldsheet models with unidexterous superfields, which satisfy the {\em\/partial\/} on-shell (``on half-shell'') constraints $\vd_\mm\BL=0=\vd_\pp\bY$ as part of their definition\eqs{e:L}{e:R}.
 Such superfields may be very useful in providing a complex, hierarchically nested geometric structure for the field space in worldsheet models that include them. Some of the marked features of such models are as follows:
\begin{enumerate}\itemsep=-3pt\vspace{-2mm}
 \item The field space is stratified; the strata may well have different dimensions.
 \item The strata may be specified by the auxiliary (non-dynamical) fields, such as in the phase diagram in Fig.~\ref{f:L1PhD}, or by the dynamical fields themselves, such as in \Eqs{e:NonStL}{e:UL2} and\eq{e:Lin}.
 \item The dynamics of the dynamical fields associated with each stratum will vary, thus describing different phases of the model at hand. The stratified space of non-dynamical fields thereby becomes the {\em\/de facto\/} phase diagram of the model at hand.
 \item The number of the effective dynamical degrees of freedom (left propagating independently upon enforcing the various constraints) may well vary from phase to phase, giving it the structure reminiscent of the ``brane-world'' scenarios.
\end{enumerate}

In addition, any model including the Lagrangian terms\eq{e:L1} also:
\begin{enumerate}\itemsep=-3pt\vspace{-2mm}
 \item provides a $(1,1|2,2)$-supersymmetric generalization of the Zeeman couplings from Ref.\cite{r6-7a};
 \item ``imports'' the external magnetic flux of Ref.\cite{r6-7a} into a worldsheet lefton superfield as $\Lb_\pp\subset\BL$, or more properly its vacuum (background) expectation value;
 \item assigns $\El,\lb_+,\bl_+\subset\BL$ as the super-partners of this ``imported'' magnetic flux, thus providing the superfield $\BL$ the physical interpretation of a magnetic super-flux.
\end{enumerate}

Finally, even with just quadratic dependence on the leftons $\BL$, Lagrangians exhibit a marked novelty as compared with the usual $(1,1|2,2)$-supersymmetric systems studied, such as Refs.\cite{rGHR,rPhases,rMS}|to name but a few earliest ones: Owing to their coupling to the {\em\/unidexterous\/} superfields $\BL$, component fields of the other superfields themselves end up being constrained dynamically, and in a left-right asymmetric fashion such as exhibited by the system\eq{e:UL2}.
 This affords a worldsheet Lagrangian framework within which to explore and generalize left-right asymmetric constructions such as the works of Ref.\cite{rAsymOF,rDK1}|again, to name but a few earliest ones.

While the analysis presented herein involved only the lefton superfields $\BL$, analogous results are obtained by using {\em\/righton\/} superfields $\bY$ in place of $\BL$, and it is evident that the two kinds of unidexterous superfields may well also be used in conjunction. This provides for a wide variety of left-right $(\vd_\mm\iff\vd_\pp)$ asymmetric constructions, which we hope to explore in more detail and with more direct application under separate cover.

\bigskip\paragraph{Acknowledgments:}
 This work was supported by the Department of Energy through the grant DE-FG02-94ER-40854. TH wishes to thank for the recurring hospitality and resources provided by the Physics Department of the Faculty of Natural Sciences of the University of Novi Sad, Serbia, where part of this work was completed.

\bigskip\bigskip
\appendix
\section{Computational Details}
 \label{s:D}
\paragraph{Berezin integrals as superderivatives:}
A single fermionic integration, the so-called `Berezin integration', is defined to equal the partial derivative and is equivalent to a superderivative:
\begin{equation}
 \int\rd\vs^+(\cdots):= [\vd_+\6(\cdots)]\big|
 = [(D_+-i\vsb^+\vd_\pp)\6(\cdots)]\big|
 ~\simeq~ [D_+\6(\cdots)]\big|\pmod{\vd_\pp\6(\cdots)}. \label{eD+}
\end{equation}
We therefore adopt the fully antisymmetrized 4-fold superderivative for the 4-fold integration:
\begin{align}
 \int\rd^4\vs(\cdots)&:=\inv{4!}
   \Big(\big\{[\Db_+,D_+]\,,\,[\Db_-,D_-]\big\}
        +\big\{[D_+,D_-]\,,\,[\Db_-,\Db_+]\big\}
         +\big\{[D_+,\Db_-]\,,\,[\Db_+,D_-]\big\}\Big)(\cdots)\Big|,\nn\\
 &~\id D^4(\cdots)\big|. \label{eD4}
\end{align}
Owing to~\eq{eSusyD},
\begin{align}
 \big\{[\,D_+\,,\,D_-\,]\,,\,[\,\Db_-\,,\,\Db_+\,]\big\}
 &= \big\{[\,\Db_+\,,\,D_+\,]\,,\,[\,\Db_-\,,\,D_-\,]\big\}-2\vd_\pp\,\vd_\mm,
  \label{eDDDbDb}\\
 \big\{[\,D_+\,,\,\Db_-\,]\,,\,[\,\Db_+\,,\,D_-\,]\big\}
 &= \big\{[\,\Db_+\,,\,D_+\,]\,,\,[\,\Db_-\,,\,D_-\,]\big\}+2\vd_\pp\,\vd_\mm,
  \label{eDDbDbD}
\end{align}
so that
\begin{subequations}
 \label{e:D4}
\begin{align}
 D^4
 &=\inv8\big\{[\,\Db_+\,,\,D_+\,]\,,\,[\,\Db_-\,,\,D_-\,]\big\}, \label{eD4u}\\
 &=\inv8\big\{[\,D_+\,,\,D_-\,]\,,\,[\,\Db_-\,,\,\Db_+\,]\big\}
   +\inv4\vd_\pp\,\vd_\mm, \label{eD4c}\\
 &=\inv8\big\{[\,D_+\,,\,\Db_-\,]\,,\,[\,\Db_+\,,\,D_-\,]\big\}
   -\inv4\vd_\pp\,\vd_\mm \label{eD4t}
\end{align}
\end{subequations}
are all equivalent, up to worldsheet derivatives, whence the simple formula\eq{e:DI}.

\paragraph{Lagrangian:}
Given any scalar function $\BK=\BK(\BL;\bX)$, we expand:
\begin{subequations}
 \label{e:D4K}
{\small
\begin{alignat}{3}
 \SL_\BL&\Defl\inv4[\Db_+,D_+][\Db_-,D_-]\,\BK(\BL;\bX)\big|
  =\inv2[\Db_+,D_+]\BK^\pp\big|,\\[1mm]
 &=\inv2\big\{\big([\Db_+,D_+]\BL^a\big)\BK^\pp_{\[a}
              +\big[(\Db_+\BL^a),(D_+\BL^b)\big]\BK^\pp_{\[ab}\nn\\
 &\mkern24mu
              +2(\Db_+\BL^a)(\ha{D}_+\BK^\pp_{\[a})
              -2(D_+\BL^a)(\ha{\Db}_+\BK^\pp_{\[a})
              +\big([\ha{\Db}_+,\ha{D}_+]\BK^\pp\big)\big\}\big|,\\[-6mm]
\intertext{where\vspace{-3mm}}
 \BK^\pp&\Defl\inv2[\Db_-,D_-]\BK,
\end{alignat}}%
\end{subequations}
and where $(\Dh_+\BL^a)=0=(\Dbh_+\BL^a)$; that is, $\Dbh_+,\Dh_+$ act on all superfields except the $\BL^a$. This results in \Eq{e:LagL}.

\bigskip\bigskip
\def\rasp{\leavevmode\raise.45ex\hbox{$\rhook$}}

\end{document}